\documentclass[aps,pre,twocolumn,superscriptaddress,secnumarabic,showpacs,floatfix,amsmath,amssymb]{revtex4}
\usepackage{latexsym}
\usepackage{amsmath}
\usepackage{amssymb}
\usepackage{amsfonts}
\usepackage{graphicx}

\begin{document}

\title{Acoustic emission before avalanches in granular media.}
\author{Vincent Gibiat}
\affiliation{Universit\'{e} Paul Sabatier, PHASE, Toulouse. France.}

\author{Eric Plaza}
\email{eplaza@inzit.gob.ve}
\affiliation{Instituto Zuliano de Investigaciones Tecnol\'{o}gicas, INZIT. Maracaibo. Venezuela.}

\author{Pierre De Guibert}
\affiliation{Universite Paul Sabatier, PHASE, Toulouse, France.}

\date{\today}
\begin{abstract}
Avalanches of granular media are mainly characterized by the observation and the measurement of the main angles of avalanche corresponding first to the movement of a group beads and to the whole movement of a great part of the grains.  These characteristics do not give any information about the rearrangements of the grains inside the granular beads layer.  As any movement of a grain produces a deformation of the structure, it is quite normal to expect a sound that will propagate inside the granular medium. We present an acoustical experimental study of the of avalanches precursors signals in spherical granular beads glass and silica aerogels powder (size of grains less than 80 micrometers). Movements inside the material produce sound pulses, these acoustic emissions have been recorded with two piezoelectric transducers placed on the lower part of the material layer. Our results show clearly signals before any movement on the upper part of the bead's layer: recursive precursors in glass beads and non-recursive precursors in aerogels, for angles less than the first avalanche angle. Our methodology allow us to obtaining information of all internal movement in the granular material including avalanches, precursors, and small arrangements. 
\end{abstract}

\maketitle
\section{Introduction to the acoustic propagation in avalanches}

The great interest of the study of avalanches is oriented to the prevention of natural disasters  and geophysical problems.  Experimental works made by \cite{11,66} define an avalanche as a self organized critical phenomenon. When an avalanche is triggered by the inclination of the pile up to the threshold of instability, the free surface of the pile defines a critical angle, frequently called maximum stability angle $\theta_m$. During the slide, a decrease of the slope of the free surface is observed until the avalanche stops. At this point the free surface defines a second critical angle: The repose angle $\theta_r$. While the packing is being tilted, one can observe many superficial rearrangements, for angles just below $\theta_m$. So an avalanche occurs when the critical angle of the bulk of material $\theta_m$ is passed by his own tilt and the pile "trying to organize itself" slips material to keep the angle of repose $\theta_r$. Recent works show evidence of events that occur before large avalanches\cite{11,22,33,55}, these events could be interpreted as avalanche predictors, they occur with angular regularity during an avalanche process.   This angular regularity opens the prediction possibility of  larger intensity events like larger avalanches.
  
In \cite{11} Nerone et al.  studied the phenomenon of avalanches through an optical system recording a video of glass beads movements enclosed in a box when they change the angle of tilt of his assembly. The system counts the number of spheres that move when they change the angle of the box in one axis, achieving and identifying precursors. This work defines the events that occur in a avalanche as: Small arrangements: small clusters formed on the free surface areas of the bulk when the small arrangements start, some of these beads are moved to other near place, and always this relation destabilizes the others beads. These events occur throughout all the experiment and anywhere on the surface.  Precursors: These events involve all beads on the surface. It is noted that the beads are uniform distributed in the free surface because the precursors are not localized events. In this case, all beads can be simultaneously moving mostly in one step, They occur for angles $\theta>17^{\circ}$ and have a regularity of $\Delta\theta\sim1.8^{\circ}$. Avalanches: Events that begin as small perturbations located anywhere in the free surface with rapid growth and that will be propagated in some lower layers.  Avalanches start at $\theta>27^{\circ}$. \cite{33} Study the precursors by a non-linear acoustic probing methodology which is sensitive to the state of weakest inter-grain contacts. For his experiment they emit a signal by an acoustical transducer. This signal travels inside of the beads bulk in the movement box process in order to obtaining in other transducers the result signals with the information of all mixed systems. The measures are numerically handled in order to separate the bead's movement signals and the noise, obtaining by this methodology the arrangement signals. This technique provides information of the state of weak contacts and only the small arrangements. 

We propose a different system to those used it by \cite{11,33} by an acoustical methodology to detect all the information of the beads movement, in a more simple way. An acoustical study of the pre-avalanche dynamics in a 3D granular packing in two different granular media enclosed in a box. We focus in the behavior of events that precede the avalanches and a non noise system. Any mechanical variation or disarrangement of the initial state imposes constraints; these constraints have a vibrating signature that can be detected as an acoustical signal and can be recorded. The principal idea of our set up is to obtain information of the internal beads movement and not only the superficial movements. For this, we design a mechanical system formed by a transparent plastic box containing the granular material. In the bottom part of the box 2 transducers make the signals registration. For triggering the avalanches the box changes softly the angle of tilt in order to register the signals produced by the material movement. We design  experiments in two different granular materials in order to compare the avalanche phenomenon.  

The first experiment was carried out with glass beads of 3mm of diameter. When we varying the inclination angle, beads start to move and this movement will produce a signal product of contact and friction between the beads of glass that move inside the material. The first movements in this assembly may be seen as precursors of the avalanche studied by  \cite{11, 22, 33, 44, 55}. These precursors will be produced by structural instability's inside of  the arrangement of the beads by the inclination of it.  

The second experiment was performed with silica aerogel. Snow is very difficult to manage, for temperature problems and non reproducibility factors; the question has been to identify a material with similar snow characteristic, especially with similar cohesion properties. It was already demonstrated by Ishida \cite{99} that snow can be treated explicitly as a granular-porous material. Previous studies on granular silica aerogel \cite{101} have provided extensive information on the characteristics in very low density and highly porosity materials. When the size of beads is less than 40 micrometers in granular materials, \cite{105} gravity force is less intense than the electrostatic. As a result, in aerogel powders (diameter size between 40-80 micrometers), electrostatic and gravity forces compete each other. Following this property, the cohesion forces of snow can be replaced by the electrostatic forces in aerogels, emulating the snow behavior. This powder gives the same "natural" features; fractures and block avalanches producing a perfectly mimicking like in a snow at a very small scale. The speed of sound is very low in silica aerogel powder (around 60 $m/s$ in comparison to 250 $m/s$ for snow) \cite{104} however the same feature is also expected for the fractures propagation acoustic emission signals. In both cases we make the signals frequency analysis. This analysis provides a way to view the avalanches process as a statistical phenomenon that gives information of the statistical periodicity during the avalanches process. 

\section{Experimental setup}
Our study tries to focus in conditions for measures without external noise in order to obtain clear acoustic signals.  The assembly took place in a transparent plastic box of Methacrylate of 4cm Height,  40cm of long and 40cm wide (see FIG 2). The height of the layer was 3cm for glass beads and 2cm for the silica aerogel powder. The box changes its tilt angle using a mechanical system articulated on its "upper" side, while the lower side is mobile and the whole box can rotate around this side axis. The rotation is controlled through an inflated balloon. When the balloon is slowly deflated the box is rotating slowly and its position and  velocity can be manipulated by stopping or regulating the quantity of air that's let out in the deflating process with a valve.  The pile is softly tilted to avoid noise disturbance by the mechanical system, up to the threshold of instability where the avalanches are triggered. On the box bottom plate two Piezo-electric transducers of resonant cells (resonance frequency around 3 kHz) make the acquisition. The acoustical signal has been recorded on both transducers at a 12 bits quantization oscilloscope for a duration of 50sec, 400Hz sampling frequency and 20k samples for the silica powder, and 200sec, 1kHz of sampling frequency and 200k samples in glass beads. We can see a Description of the experimental set-up in FIG (1):

\begin{figure}
\includegraphics[scale=0.3]{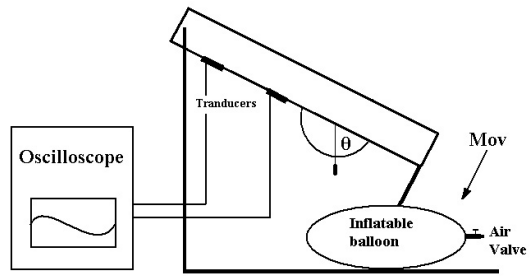}
\centering
\caption {Scheme for the miniature avalanches experiment. The angle of the box  $\theta$  will vary softly when the ball is deflated. The ball touch the box through a metal rod in order to deflate the ball avoiding mechanical noise. The velocity of the downing process is controlled by the air valve}
\end{figure}

In the lowering process the velocity of angle variation is roughly linear because the angular velocity is slow. The angle variation can be directly related to the box lowering time in order to interpret the angular variation in the graphics.

This simple system allows repeatable experiments with the advantage of  minimal noise. In order to minimize the mechanical wave conduction between the set up and the laboratory floor, we place the experimental set up over a foam surface. FIG (2) represent the box rotation without any material inside as a test of noise in the lab. The signal level was around 0.00079mV while the smallest recorded signals corresponding to an avalanche have amplitude of around 0.02781mV and the regular signals have amplitude between 0.5mV. In the FFT graphics for frequencies domain, we report  two specific frequencies in 50Hz and 232 Hz. These frequencies are present in all the experiments and correspond to the lab noise produced by a motor noise in the nearest of laboratory building.

\begin{widetext}

\begin{figure}
\includegraphics[width=0.8\textwidth]{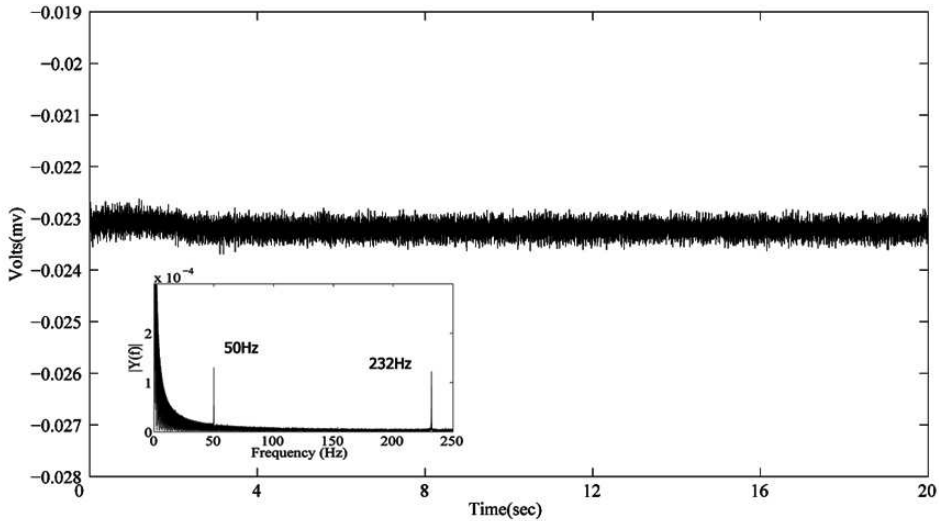}
\centering
\caption{ System downing measures without material. Middle Transducer 500Hz 10k samples 20sec. And FFT}
\end{figure}

\begin{figure}
\includegraphics[width=0.8\textwidth]{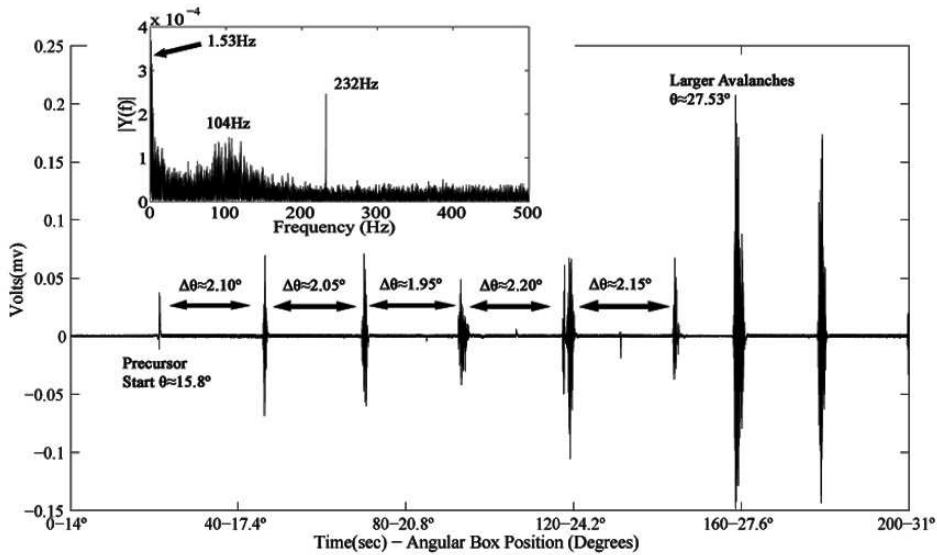}
\centering
\caption{Beads of glass 3mm. Middle Transducer 1KHz 200k samples 200sec and FFT}
\end{figure}

\begin{figure}
\includegraphics[width=0.8\textwidth]{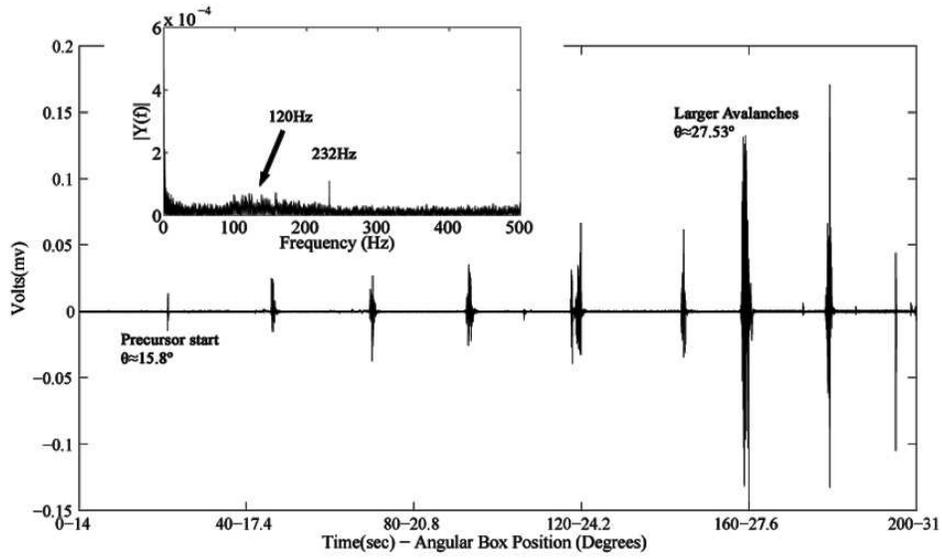}
\centering
\caption{Beads of glass 3mm. Upper Transducer 1KHz 200k samples 200sec and FFT}
\end{figure}

\begin{figure}
\includegraphics[width=0.7\textwidth]{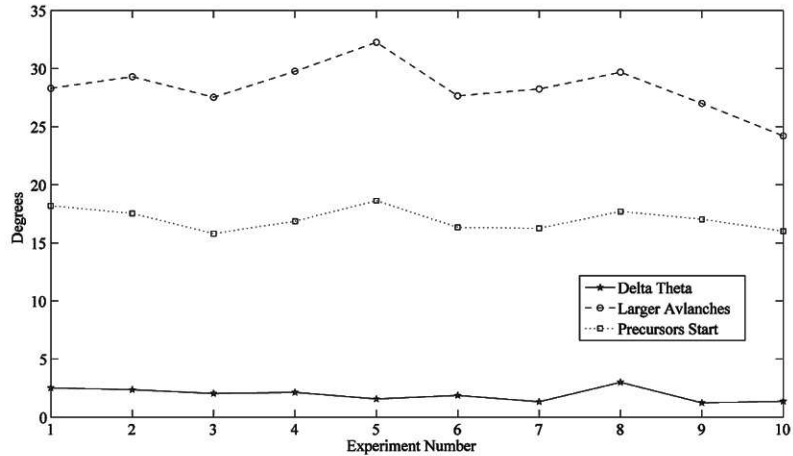}
\centering
\caption{Beads of glass 3mm. Upper Transducer 1KHz, 200k samples 200sec. Precursor Start Point, Larger avalanches and $\Delta\theta$ value for 10 experiments.}
\end{figure}

\begin{table}
\begin{center}
\begin{tabular}{|c|cccccccccc|c|c|c|}
	\hline
Experiment Number & $1$ & $2$ & $3$ & $4$ & $5$ & $6$ & $7$ & $8$ & $9$ & $10$ &  $Mean$ &  $SD$ \\
	\hline

Precursors Start     & 18.3 & 17.5 & 15.8 & 16.9 & 18.6 & 16.3 & 16.3 & 17.7 & 17.0 & 16.01 &  $17.04$ &  $0.326$\\

Larger avalanches    & 28.3 & 29.3 & 27.5 & 29.8 & 32.3 & 27.7 & 28.2 & 29.7 & 27.0 & 24.3 &  $28.41$ &  $1.299$\\

Number of precursors & 4 & 5 & 5 & 6 & 8 & 6 & 9 & 4 & 8 & 6 &  $6.10$ &  $0.031$\\

$\Delta\theta_n$       & 2.5 & 2.3 & 2.0 & 2.1 & 1.6 & 1.9 & 1.3 & 3.0 & 1.2 & 1.4 &  $1.94$ &  $0.182$\\

Standard deviation   & 0.22 & 0.05 & 0.02 & 0.2 & 0.28 & 0.22 & 0.22 & 0.36 & 0.21 & 0.24 &  $-$ & $-$\\

	\hline
\end{tabular}
\caption{3mm Glass Beads information for each experiment.}
\end{center}
\end{table}

\begin{figure}
\includegraphics[width=0.8\textwidth]{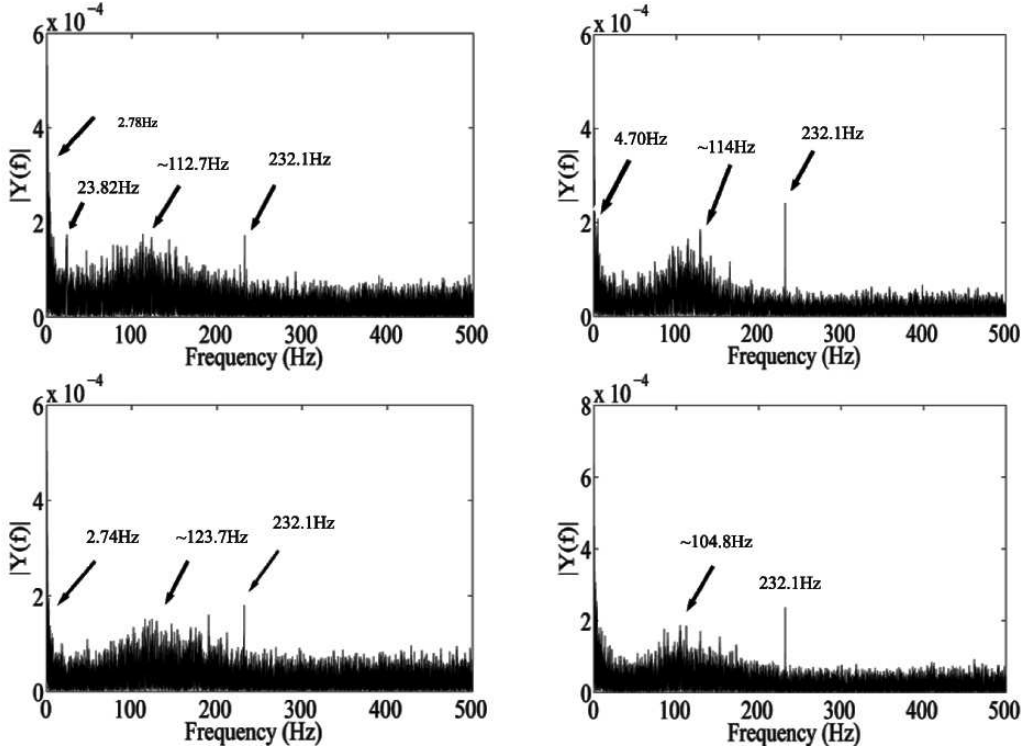}
\centering
\caption{ 4 Glass beads experiments FFT. Middle transducer}
\end{figure}

\begin{figure}
\includegraphics[width=0.8\textwidth]{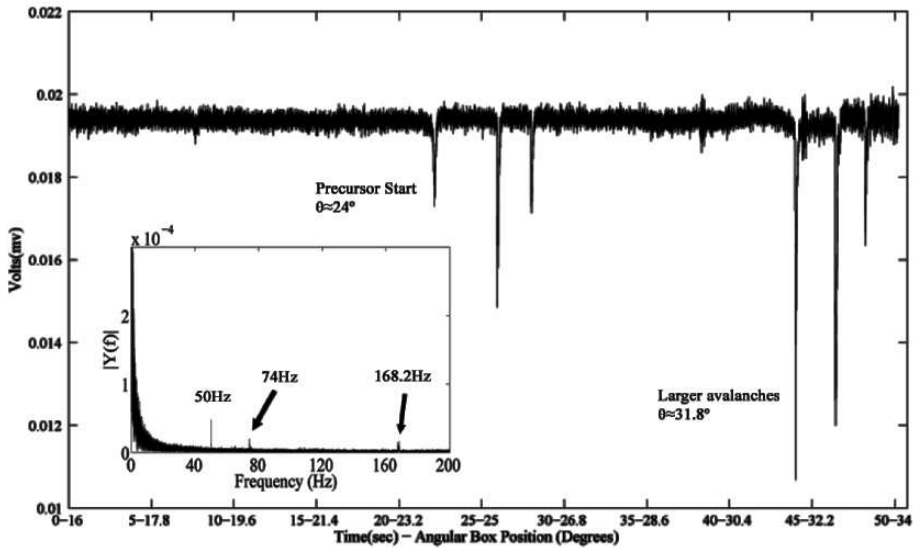}
\centering
\caption{ Aerogel. Middle Transducer. 400Hz 20k samples 50sec and FFT}
\end{figure}

\begin{figure}
\includegraphics[width=0.8\textwidth]{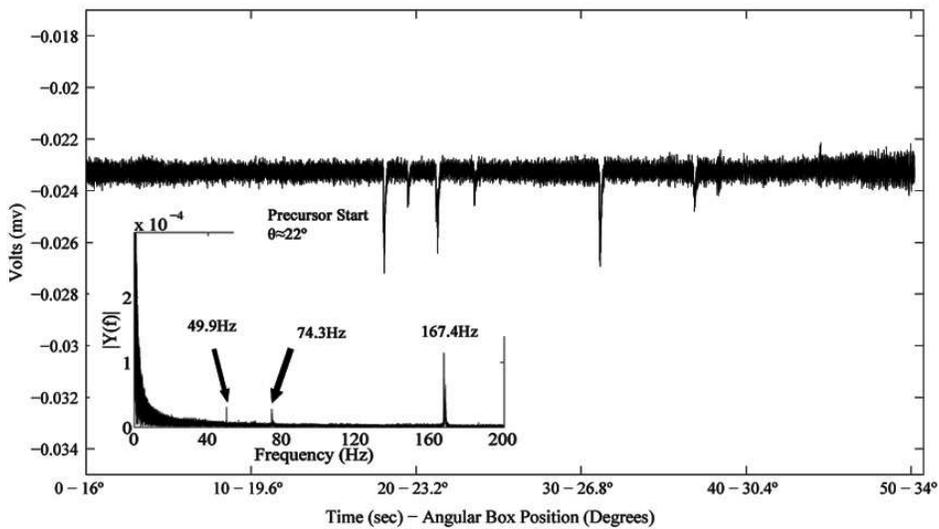}
\centering
\caption{Aerogel. Upper Transducer. 400Hz 20k samples 50sec. And  FFT}
\end{figure}

\begin{table}
\begin{center}
\begin{tabular}{|c|cccccccccc|c|c|c|}
	\hline
Experiment Number & $1$ & $2$ & $3$ & $4$ & $5$ & $6$ & $7$ & $8$ & $9$ & $10$ &  $Mean$ &  $SD$\\
	\hline

Precursors Start     & 21.6 & 24.3 & 27.1 & 25.3 & 12.6 & - & 23.97 & 23.5 & 22.2 & 24.7 & $22.80$ &  $0.63$\\

Larger avalanches   & 30 & - & - & - & - & - & 31.8 & 32.3 & - & 34.0 &  $32.02$ &  $0.99$\\

Number of precursors & 3 & 3 & 2 & 3 & 2 & - & 5 & 2 & 1 & 2 &  $2.30$ &  $0.09$\\

$\Delta\theta_n$       & 2.9 & 6.7 & 1.8 & 3.3 & 5.9 &  & 2.0 & 8.8 & - & 3.4 &  $4.35$ &  $0.34$\\

Standard deviation   & 0.1 & 0.6 & - & 2.1 & - & - & 0.6 & - & - & 1.4 &  $-$ &  $-$\\

	\hline
\end{tabular}
\caption{Aerogel information for each experiment. Middle transducer.}
\end{center}
\end{table}

\begin{table}
\begin{center}
\begin{tabular}{|c|cccccccccc|c|c|c|}
	\hline
Experiment Number & $1$ & $2$ & $3$ & $4$ & $5$ & $6$ & $7$ & $8$ & $9$ & $10$  &  $Mean$ &  $SD$\\
	\hline

Precursors Start     & - & - & - & - & 26 & 18.4 & 22.5 & 19.1 & 19.9 & 25.6 &  $21.91$ &  $1.50$\\

Larger avalanches    & - & - & - & - & 29.4 & 33.8 & - & - & 32.1 & 28.5 &  $30.95$ &  $1.23$\\

Number of precursors & 0 & 0 & 0 & 0 & 6 & 5 & 6 & 7 & 8 & 4 &  $3.6$ & $0.13$\\

$\Delta\theta_n$       & - & - & - & - & 1.4 & 4.3 & 1.5 & 2.3 & 1.5 & 1.5 &  $2.08$ &  $2.24$\\

Standard deviation   & - & - & - & - & 0.4 & 1.3 & 0.3 & 0.5 & 0.2 & 0.2 &  $-$ &  $-$\\

	\hline
\end{tabular}
\caption{Aerogel information for each experiment. Upper transducer.}
\end{center}
\end{table}

\begin{figure}
\includegraphics[width=0.8\textwidth]{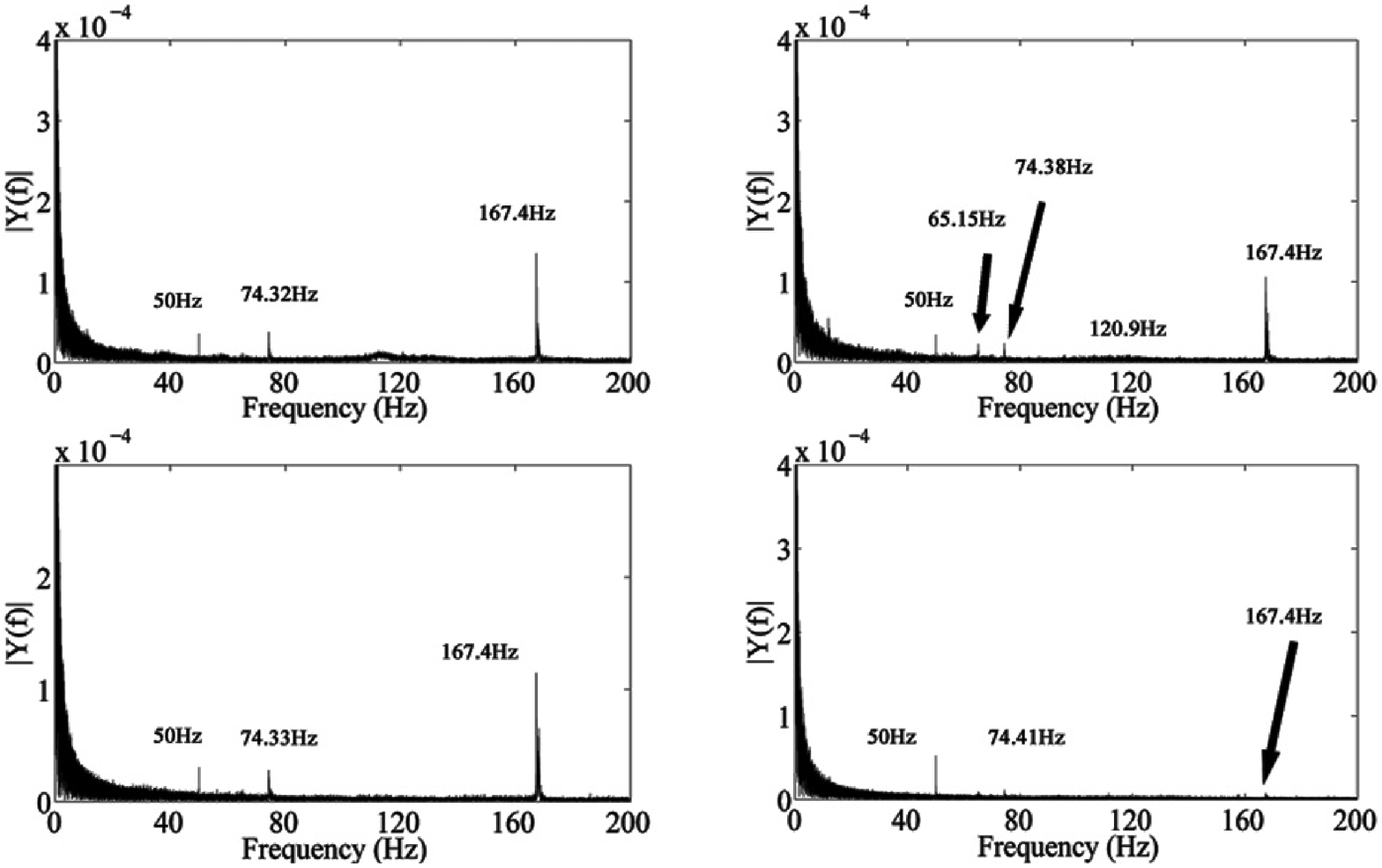}
\centering
\caption{ 4 aerogel experiments FFT. Upper transducer}
\end{figure}

\end{widetext}
We perform 10 experiments for glass beads and 10 experiments for aerogel powder. Each one under the same experimental conditions. We compute the angular velocity in the box downing process for each experiment using the acquisition time and the familiar expression of lineal  angular velocity $v=\theta/t$ in order to obtain the angular path for each case. The experimental time was assumed long for be able to approximate at a lineal angular velocity behavior. With this velocity we can compute the signals angular position for each graphic.

In the FIG (3 and 4) we present the signals obtained for the transducer located in the middle and upper part of the Plexiglas bottom box in one of the 10 experiments. The signals observed correspond to the beads movement inside the box. The amplitude of these signals is associate to the quantity of moving beads. Less signals events are associate to the precursors and more amplitude to the larger avalanches. There are a remarked different between both kinds signals in the graphics. Clear precursors events signals are observed with a angular regular interval mean of $\Delta\theta=2.09^{\circ}$. The compute of the standard deviation (SD) of these mean value give us a measure of the data dispersion and the proximity in the regularity for these events. More similar events will be produce small values in DE. In this experiment $ED=0.042$. The precursors start for an angle of $\theta=15.8^{\circ}$ and the avalanches start in $\theta=27.53^{\circ}$. 
For all the 10 experiment we present the table 1 and FIG (5). In these we can observe the behavior in 10 ten experiments in order to compare. We present the measures for each experiment for: the precursor's start angle, mean precursors interval, mean precursors interval SD, and avalanches start angle. We present the arithmetic mean for the characteristic angles for all experiment (see table I). These values present a clear tendencies in the precursors start angle $\theta=17.04^{\circ}$ with a dispersion of $SD=0.326^{\circ}$, avalanches start  $\theta=28.41^{\circ}$ with $SD=1.29^{\circ}$, the size of the precursors mean interval was $\Delta\theta=1.94^{\circ}$ with $SD=0.18^{\circ}$.  These values are agree with the Nerones work in \cite{11} where they values for the precursors start in $\theta>17^{\circ}$ large avalanches for $\theta=27.3^{\circ}$ and precursor angular interval of $\Delta\theta=1.8^{\circ}$. The small differences are maybe associate to the fact that the avalanches are not a lineal phenomenon, for that its impressive to make a more higher number of experiment in order to compare the mean behavior of it. 

The Fourier transform gives us information about the events repetition inside the signal. If we imagine one beads downing in a beads layer. This bead can collision against the others and produce a sound that can be register in a signal.  These signals will have a period associate to the time between collisions, and this period a frequency associate that can be observed in a FFT. For our case we have a lot of bead moving in the box, and the FFT spectrum gives us information of the statistical repetitive events. We present in the FIG(6),  4 experiments with the same conditions, a clear Gaussian behavior cantered in approximate 100Hz is reported. These frequencies have a period of 0.001sec. We assume this period as a statistical measure of time collisions in the bulk of material when they change his tilt. These signals correspond to those reported for Nerone in \cite{11} as small arrangements in his optical system. This acoustical signals methodology give a easier and complete way to obtain all the avalanche informations and to apply in field observations.

In the aerogel the results are different. The avalanches are block avalanches like in the snow case\cite{77}. This avalanches start by internal fractures inside the material, our propose is to detect the fracture signals as a precursors events.  Is convenient to remember that the aerogel is a damping material \cite{101}. This feature prevents the detection of signals coming from distant source locations. As a result, there are differences recorded amplitude values between the transducer located at the top and the middle. In our case we perform 10 experiments. Each one with similar characteristics, in a 50 seconds measurement interval 20k samples and 400Hz. FIG. (7 and 8) show the behavior of an amplitude and frequency of one of the 10 experiments in aerogel in the middle transducer and upper respectively. The amplitude of the signals in Figure 7 are different from those show it in FIG (8) for the reasons explained above, however, the Fourier spectrum  in both cases give similar results 167Hz and 74Hz signals. These frequencies correspond to short periods vibration caused by the attenuating properties of the material. In the FIG (9) we present the behavior of the frequencies in 4 different experiments. The 50Hz frequency is not taking in a count in the analysis because correspond to the laboratory noise as we can clearly see in the graphics in FIG 3.

As a result of the graphics observation we can affirm that does not exist angular regularity in the signals amplitude events in FIG (7 and 8). This is reaffirmed in the table II and III where the results present a higher SD values  in comparison of his own mean for the precursors interval  $\Delta\theta_n$ for each particular experiment. In many cases there are not presence of  significative amplitude signals for consider as avalanches or even precursors. However this situation only occur  in 1 of the 2 transducer (see tables II and III). If there are not  signals in one, the other yes have it. That's because there are not enough energy to propagate the acoustic wave for all the material bulk.

\section{Conclusions}
Our low noise levels system allowed us to obtain the signals that occur when the pile is tilted to the threshold of stability in beads of glass and powder of aerogel.  In the glass beads case, we reproduce the results obtained in \cite{11} using a acoustical system and comparing in 10 experiments. We processing the signals with a FFT obtaining the frequency spectrum.  The frequencies analysis show us results associated to the movement of the complete bulk of material. Obtaining by this methodology the mean of the time movement period in the glass beads or an mean of frequency movement in small arrangements. In the aerogel powder, it's the first evidence of block avalanches precursors signals. In this case we obtain signals only in the proximity of transducer because his damping characteristic. However middle intensity signals are reported before the great avalanches, showing evidence of non regular precursor. This system confirm the possibilities of use this kinds of acoustical setup in real field scenarios. This signals despite not have recursively can be take it as a predecessor of the larger avalanche events. Despite both are a granular material the behavior of avalanche granular dynamics are different. The interaction between the grains produce different phenomenon because the different individual weight for the glass beads particles and the aerogel particles. In the glass beads case (heavier material) the gravity produce over the grains a faster individualized movement, the gravity force is more significant in comparison of cohesion forces. This force act over each particles un-stabilizing the material bulk when it arise his critical angle.  This instability  produce the localized phenomenon of precursor and the larger avalanche. The precursors are caused by the dis-stabilisation of the individual grains over his neighbor. When the instability arise a more significant level, all the material slip down. In the aerogel this individual behavior is more retarded because the electrostatic force interaction inter-grain is more significant in comparison to the gravity and this act like the cohesion force. The grain contact keep without moving the aerogel bulk particles even near to critical angle, the avalanches occur only when the threshold of instability is more intense. The avalanche start in fractures, when the first particles are moved  because the effect of gravity force wining vs the cohesion force, the material start to separate this cause a chain reaction that separate a lot of near particles causing and propagating  the fracture. If this instability continue the fractured material is divided in small piece of materials. and the gravity slip down this more heaviest pieces increasing the instability and producing the avalanches.


\end{document}